\documentclass[prl,twocolumn,showpacs]{revtex4}

\usepackage{graphicx}
\usepackage{hyperref}
\usepackage{amssymb,amsmath}
\usepackage{epstopdf}

\begin{document}

\title{Quantum Decoherence of Single-Photon Counters}

\author{V. D'Auria, N. Lee, T. Amri, C. Fabre and J. Laurat}

\affiliation{Laboratoire Kastler Brossel,
Universit\'{e} Pierre et Marie Curie, Ecole Normale Sup\'{e}rieure,
CNRS, Case 74, 4 place Jussieu, 75252 Paris Cedex 05, France}

\begin{abstract}
The interaction of a quantum system with the environment leads to the so-called quantum decoherence. Beyond its fundamental significance, the understanding and the possible control of this dynamics in various scenarios is a key element for mastering quantum information processing. Here we report the quantitative probing of what can be called the quantum decoherence of detectors, a process reminiscent of the decoherence of quantum states in the presence of coupling with a reservoir. We demonstrate how the quantum features of two single-photon counters vanish under the influence of a noisy environment. We thereby experimentally witness the transition between the full-quantum operation of the measurement device to the "semi-classical regime", described by a positive Wigner function. The exact border between these two regimes is explicitely determined and measured experimentally.
\end{abstract}

\date{\today}
\maketitle

The coupling of quantum systems to an environment leads to the transition from the quantum to the classical worlds. The study of this transition has raised a great deal of work over the past decade \cite{Zurek2,Haroche}. For example, seminal experiments have allowed the controlled probing of the decoherence of quantum states, including mesoscopic superpositions in microwave cavity \cite{Brune, Haroche2}, motional states of a trapped ion \cite{Turchette}, spatially separated atomic superpositions \cite{Kokorowski}, and amplified number states \cite{Zavatta}. If decoherence lays at the heart of the foundations of quantum physics, this process is also of pratical importance as it plays a central role in quantum information processing \cite{Nielsen}. In the present work, we complement this study by adressing the effect of a decohering environment, not on a quantum state, but on the quantum capability of a measurement apparatus. 

We consider here how a noisy environment quantitatively degrades the quantum performances of optical detectors. This corresponds to many pratical situations where dark counts, additional background or any undesired emissions into the detected mode can degrade the expected performances of the measurement, which therefore limit their use in a large range of applications. Precisely assessing the quantum features of photon counting devices plays indeed an increasing role in the development of quantum technologies \cite{Hadfield}. For example,
measurement-driven information processing \cite{Knill,OBrien2}, quantum key distribution \cite{Gisin} and state engineering
\cite{Illuminati,Lvovsky,Alex1,Neergaard,Takahashi,Lvovsky2} rely more and more on such countings. The control of the non-classical features of the detectors used in these protocols is central to this endeavour.

\begin{figure*}[t]
\begin{minipage}{1.2\columnwidth}
\includegraphics[width=1\columnwidth]{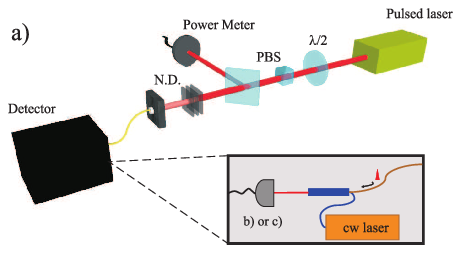}
\end{minipage}
\begin{minipage}{0.4\columnwidth}
\includegraphics[width=1\columnwidth]{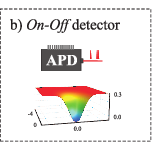}\\
\includegraphics[width=1\columnwidth]{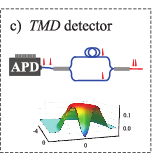}
\end{minipage}
\caption{{(color online). Experimental setup. (a) The  "black-box" detector is probed with
coherent pulses of well calibrated amplitudes coupled into a single-mode fiber. The polarizing beam-splitter (PBS) provides a first attenuation
and a small ratio of the beam is then tapped and measured with a powermeter. A set of neutral density filters (ND) gives a subsequent strong
attenuation ($\simeq 10^7$) to reach the regime of few photons per pulse.  In order to investigate the effect induced by a noisy environment, different background noise levels are simulated via a continuous-wave laser mixed with the probe pulses on a 50/50 fiber
beam-splitter. Two devices are characterized: (b) a conventional single-photon counting module (APD) providing an on-off outcome and (c) a single-loop time-multiplexed detector (TMD) with photon-number resolution ability. The 3d-plots correspond to the experimental Wigner function
associated with the one-click outcome for each device without added noise.}}
\end{figure*}

In order to follow the decoherence of detectors exposed to a controlled noisy environment, we experimentally determine their positive operator valued measure (POVM) \{$\hat{\Pi}_{n}$\}  \cite{Soto,Fiurasek,D'Ariano1} and assess their non-classical features. For phase-insensitive measurements, as it is the case for the single-photon detectors considered here, the outcome labelled by '$n$' corresponds
to '$n$ clicks'. In the ideal case of photon counters with unity quantum efficiency, photon number resolution and no dark counts, the POVM
operator associated to the measurement '$n$' is the projector $|n\rangle\langle n|$ in the number basis. Real devices differ from this
case and the operator $\hat{\Pi}_{n}$ can be written as a sum of projectors:
\begin{equation}
\hat{\Pi}_{n}=\sum_{l=0}^{\infty} r_{l,n}(\eta,\nu) |l\rangle\langle l| . \label{POVM}
\end{equation}
where the coefficients $ r_{l,n}\geq0$ depend on the detector quantum efficiency $\eta$ and  the mean number $\nu$ of noise counts in the
detection windows. Detector tomography leads to the experimental determination of all the coefficients $\{r_{l,n}(\eta,\nu)\}$ without any a
priori knowledge on the device \cite{Lundeen, Ronge}. The reconstructed operator can then be represented by a quasi-probability distribution, namely its Wigner
function. This distribution can take negative values, which prevents it to be interpreted as a regular probability
distribution. The existence of such negative values is therefore a strong signature of the full quantum character of the
measurement device under study. In addition, it can be shown that the non-classical properties of a measurement performed by a detector can be associated with the non-classical properties of the state retrodicted from its response \cite{Taoufik}. The density matrix of the retrodicted state is given by the normalized POVM. The negativity of the detector Wigner function can thus be associated with the negativity of its retrodicted state.

Our work was carried out on two different detectors, a single-photon counting module based on an avalanche photodiode and a two-channel time-multiplexed detector with photon-number resolution ability \cite{Achilles}. We determine the Wigner functions associated with the two dectectors under study and consider in particular their negativity at the origin
of coordinates for different amounts of added noise. In analogy with decoherence studies performed on quantum states, we will
be able to precisely witness the transition of a detector from the full quantum to the semi-classical domain, where quantum fluctuations can be seen as arising from unmastered fluctuations of the electromagnetic field. We will finally illustrate how the detector decoherence manifests itself when the detector is used to herald the preparation of a target state, a paradigm for quantum state engineering.

Figure 1a shows a schematic of the experimental setup. Detectors are probed  with a tomographically complete set of  coherent states $|\alpha\rangle$ of different amplitudes obtained from a pulsed laser source at 795 nm with a repetition rate set to 1.187$\pm$0.001 MHz (Mira900 with Pulse Picker 9200, Coherent). The average photon
number per pulse can be varied from 0 to 100, with residual laser intensity fluctuations between subsequent pulses measured to be 5\% (peak to peak).The two detectors under study are contained in the black box of Fig. 1a, with details in insets 1b and 1c. The first device is an avalanche photodiode (APD, Perkin Elmer SPCM-AQR-13), which corresponds to an "on/off detector" as it provides only two possible
outcomes corresponding to one-click and no-click. This binary  response can only distinguish between zero photons and "one or more". The second
detector is home-made and designed following the time-multiplexed detection (TMD) scheme of Ref. \cite{Achilles} consisting in splitting the
light in two different time bins and detecting it with an APD. This single-loop TMD realizes
a simple version of a photon-number resolving detector as it provides three possible responses: no-click, one-click and two-clicks. In the ideal case, a single photon will provide one-click while two impinging photons have 50\% probability to fill both time bins and thus leading to a two-clicks outcome. In order to directly compare the two devices, their total quantum efficiencies have been both set to $0.28\pm0.02$. This value takes into account all the losses and the intrinsic quantum efficiency of the APD. The coupling into the fiber before the black-box is part of the probe calibration and does not contribute to this efficiency. The detection window is set to 40 ns long.

For both detectors, we collect the tomographic data consisting of the different outcome  statistics as a function of the average photon number
per pulse. The POVM density matrices are then reconstructed by a Maximum-likelihood algorithm  \cite{Fiurasek} and the associated Wigner functions are finally obtained by the sum of the Wigner functions of projectors  $|n\rangle\langle n|$, each weighted by the coefficients $ r_{l,n}(\eta,\nu)$ \cite{Mateo}. As shown in Fig. 1b and 1c, the difference between the two detectors
can be clearly seen on the general shape of their associated Wigner function.  As we will see more precisely later, the value at the origin is negative for both detectors, which is a signature of the
full quantum character of the measurement. Our work aims at following the evolution of this negativity under the influence of noise.

Experimentally, we investigate such an effect by adding an additional excitation channel. As sketched on Fig. 1a, the
background noise is controlled by injecting in the same path than the probe light a continuous-wave coherent beam at 1064 nm (Diabolo,
Innolight).  By adjusting the cw laser power, we simulate different noise levels with Poissonian statistics. This statistics is of particular interest as it simulates many practical cases for single-photon counters. Four different values are used
for the mean number of dark counts $\nu=\{0, 0.03, 0.08, 0.18\}$. The Wigner functions for the one-click POVM, $W_{on}$ for the APD and $W_{1}$ for the
TMD, are displayed in Fig. 2. Due to the absence of phase dependence, only cross-sections are given.

When the noise increases, the negativity of the Wigner functions is gradually  reduced, and finally disappears. The insets
of Fig. 2 provide the evolutions of the negativity at the origin as functions of the added noise. The observed transition from negative to positive Wigner functions witnesses here in a quantitative way the degradation of detector performances with a noisy environment. Moreover, we experimentally identify the exact border between two
different regimes for the detectors under study as $\nu\sim\eta/2$. The lower is the efficiency of the counter, the more stringent is the limit on the noise level.

These results are in agreement with the theoretical model that we outline now. For phase-insensitive detectors, the effect of Poissonian dark counts can be included in the theoretical description of POVM operators \cite{Barnett}. The detector not being able to discriminate between dark counts and photoelectrons, the probability for registering '$n$' clicks, i.e. the expectation value of $\hat{\Pi}_{n}$, is indeed the discrete convolution of the dark count probability distribution, given by $\nu^n e^{-\nu}/n!$, and the probability of '$n$' clicks in the absence of noise ($\nu$=0). By including the noise into the model of Ref. \cite{TMD}, we derived the coefficients $r_{l,off}(\eta,\nu)=e^{-\nu}(1-\eta)^{l}$ for $\Pi_{off}$ of the APD and 

$r_{l,1}(\eta,\nu)=2 e^{-\nu/2}(1-\eta)^{l}(-e^{-\nu/2}+(1+\frac{\eta}{2(1-\eta)})^l)$
for $\Pi_{1}$ of the TMD detector. The probabilities for the photon to be sent toward the long and the short
path of the TMD were assumed to be equal. Correspondingly, the value at the origin of the Wigner function associated to  $\hat{\Pi}_{on}$ is:
\begin{equation}
W_{on}(0,0)=\frac{1}{\pi}\left(1-\frac{ e^{-\nu}}{1-\eta/2}\right)\nonumber
\end{equation}
while for $\hat{\Pi}_{1}$ it can be read as:

\begin{equation}
W_{1}(0,0)=\frac{4}{\pi} e^{-\nu}\left(\frac{ e^{\nu/2}}{2-\eta/2}-\frac{1}{2-\eta}\right). \nonumber
\end{equation}
Given a quantum efficiency $\eta$,  the negativity disappears  for $\nu$ = $-\ln(1-\eta/2)$ for the APD and $-2\ln(1-\frac{\eta}{4-\eta})$ for the TMD. In both cases, for the quantum efficiency considered here, these values can be approximated by the simple formula $\nu\sim\eta/2$, as experimentally shown.

\begin{figure}[t]
\includegraphics[width=.85\columnwidth]{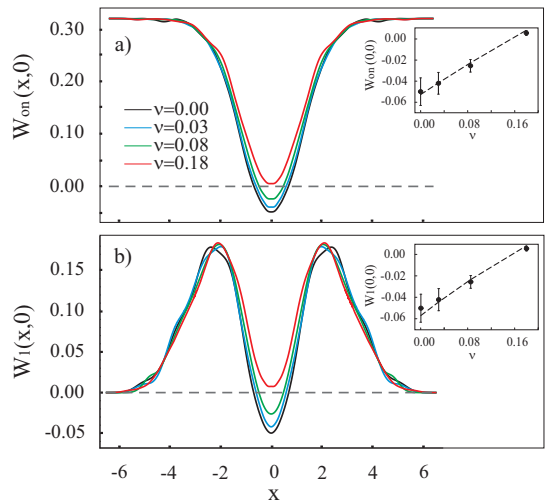}\label{sections}
\caption{(color online). Noise-induced quantum decoherence of the counters. The cross-sections of the Wigner function associated to the one-click response are displayed for (a) the APD and (b) the TMD for different values of added-noise given by the mean number $\nu$ of dark counts in the detection window. Insets give the evolution of the negativity at the origin and dotted lines correspond to theoretical predictions. Error bars are obtained by combining the following uncertainties. The numerical determination of $r_{l,n}(\eta,\nu)$ critically depends on the number $\emph{L}$ of projectors taken in the development of Eq. $\ref{POVM}$. To evaluate the uncertainty due to this truncation, the ML algorithm has been run for different values of $\emph{L}$. Another error source is the uncertainty on the average photon number of the probe states. This contribution has been checked by replacing each experimental $|\alpha_{j}|^{2}$ with a random value sorted by a Gaussian distribution centered over $|\alpha_{j}|^{2}$ and with a variance $0.025|\alpha_{j}|^{2}$ corresponding to the 5$\%$ intensity fluctuations, and by performing different tomographic runs.}
\end{figure}

Besides its fundamental interest in understanding the quantum properties of optical detectors, our  investigation has a direct impact on the
design of quantum information protocols and more precisely on the engineering of the quantum states of travelling optical fields
using a conditional measurement. A general preparation strategy consists indeed in measuring one mode of an entangled state, which results in
projecting the other mode according to this measurement \cite{Paris}. 

Given the Wigner function of the bipartite ressource $W_{ab}$,
the one of the conditional state $\hat{\rho}_c$ obtained when the measurement performed on mode a leads to the outcome '$n$' can be written in the most general way as:
\begin{equation}
W_c(x,y)=\frac{\int\,dx'dy'\,W_{ab}\left(x,y,x',y'\right)W_n\left(x',y'\right)}{\int\,dxdydx'dy'\,W_{ab}\left(x,y,x',y'\right)W_{n}\left(x',y'\right)}
\label{output}
\end{equation}
where $W_n$ is the Wigner function associated to the POVM $\hat{\Pi}_n$ and the denominator is a normalization constant. From this expression, it follows that the  preparation of a quantum state with negative
Wigner function does require a heralding detector with negative Wigner function if the entangled ressource has a positive one. The decoherence of the optical detector used in the
conditional measurement will thus translate into the decoherence of the prepared state. 

We illustrate this effect by considering as a ressource the correlated photon-pairs produced by two-mode spontaneous parametric down-conversion \cite{Illuminati}. Usually, this state is parametrized by a coefficient $\lambda$, which varies between 0  and 1 and is experimentally related to the intensity gain in the down-conversion process by the relation $\lambda^2=1-1/G$. Let us note that, in the limit of high gain ($\lambda \rightarrow 1$), which corresponds to vanishing widths of the two-mode Gaussian Wigner function $W_{ab}$, the prepared state reduces to the normalized POVM: its quantum properties are the ones of the detector. 

Figure 3 gives the theoretical Wigner function of the state obtained for a heralding measurement performed  with the APD previously characterized on a
numerically simulated entangled ressource defined by $\lambda=0.6$, a typical experimental value. As it can be seen, the quantum decoherence
transition of the detector directly translates into such transition for the engineered state. As a result, we observe a gradual transition
between a state with negative Wigner function to a state with a positive Wigner function approaching a gaussian shape and corresponding to the
classical thermal state generated by spontaneous parametric down-conversion.

\begin{figure}[t]
\includegraphics[width=.85\columnwidth]{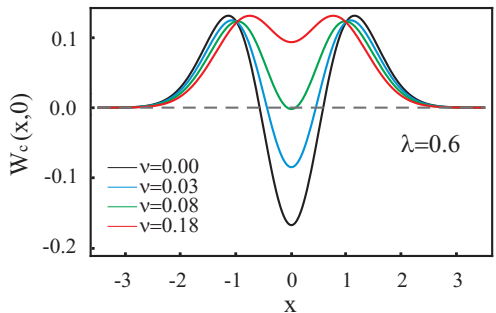}\label{cond}\vspace{-0.5cm}
\caption{(color online). Effect of the heralding detector decoherence on quantum state engineering. Given the experimental POVM of the APD, we determine the state prepared in a conditional scheme by simulating the entangled ressource. The plot gives the cross-sections of the Wigner function, $W_c(x,0)$, of the engineered state $\hat{\rho}_c$ for $\lambda=0.6$ and for the different levels of noise for which the POVM has been previously reconstructed.}
\end{figure}

In conclusion, we have adressed the quantum decoherence of optical measurement devices by explicitly following this process for single-photon counters. Our investigation has thereby provided a quantitative witness of the evolution of the behavior of a detector under external parameters which drive the
device away from the full quantum regime. For the typical single-photon counters used here, the border is given by $\nu\sim\eta/2$, as experimentally measured.  Beside being of fundamental interest for understanding measurements in quantum
physics, our study also has practical implications as it directly applies to quantum state preparation, as illustrated, or more generally to  any
measurement-driven processes. 

\acknowledgements 
We acknowledge helpful discussions with S. Olivares. This work has been supported by the  BQR from Universit\'{e} P. et M. Curie. V.D'A. acknowledges the financial support from the EC under the Marie Curie Programme and N.L. from the G-COE program commissioned by the MEXT of Japan. N.L. is with the Department of Applied Physics, The University of Tokyo.

\end{document}